\documentclass[%
reprint,
superscriptaddress,
amsmath,amssymb,
aip,
floatfix,
]{revtex4-1}

\pdfoutput=1

\usepackage{color}
\usepackage{xcolor}
\usepackage{siunitx}
\DeclareSIUnit \px {px}
\usepackage{float}
\usepackage{xspace}
\usepackage{natbib}

\usepackage{graphicx}
\usepackage{dcolumn}
\usepackage{bm}
\usepackage{color}

\newcommand{\singlecolumn}{85mm}
\newcommand{\doublecolumn}{170mm}

\newcommand{\Bone}{$\textrm{B}_{\textrm{1}}$}
\newcommand{\Btwo}{$\textrm{B}_{\textrm{2}}$}
\newcommand{\Bthree}{$\textrm{B}_{\textrm{3}}$}
\newcommand{\Bfour}{$\textrm{B}_{\textrm{4}}$}

\newcommand{\SupVoltage}{I}

\newcommand{\SupLoadingMap}{II}

\newcommand{\SupManipSubPart}{III}

\begin{document}

\title{Controlled quantum dot array segmentation via a highly tunable interdot tunnel coupling}

\author{Martin Nurizzo}
\affiliation{Univ. Grenoble Alpes, CNRS, Grenoble INP, Institut N\'eel, F-38000 Grenoble, France\looseness=-1}%

\author{Baptiste Jadot}
\affiliation{Univ. Grenoble Alpes, CEA, Leti, F-38000 Grenoble, France}%

\author{Pierre-Andr\'e Mortemousque}
\affiliation{Univ. Grenoble Alpes, CEA, Leti, F-38000 Grenoble, France}%

\author{Vivien Thiney}
\affiliation{Univ. Grenoble Alpes, CEA, Leti, F-38000 Grenoble, France}%

\author{Emmanuel Chanrion}
\affiliation{Univ. Grenoble Alpes, CNRS, Grenoble INP, Institut N\'eel, F-38000 Grenoble, France\looseness=-1}%

\author{Matthieu Dartiailh}
\affiliation{Univ. Grenoble Alpes, CNRS, Grenoble INP, Institut N\'eel, F-38000 Grenoble, France\looseness=-1}%

\author{Arne Ludwig}
\affiliation{Lehrstuhl f{\"u}r Angewandte Festk{\"o}rperphysik, Ruhr-Universit{\"a}t Bochum, Universit{\"a}tsstra{\ss}e 150, D-44780 Bochum, Germany}%

\author{Andreas D. Wieck}
\affiliation{Lehrstuhl f{\"u}r Angewandte Festk{\"o}rperphysik, Ruhr-Universit{\"a}t Bochum, Universit{\"a}tsstra{\ss}e 150, D-44780 Bochum, Germany}%

\author{Christopher B{\"a}uerle}
\affiliation{Univ. Grenoble Alpes, CNRS, Grenoble INP, Institut N\'eel, F-38000 Grenoble, France\looseness=-1}%

\author{Matias Urdampilleta}
\affiliation{Univ. Grenoble Alpes, CNRS, Grenoble INP, Institut N\'eel, F-38000 Grenoble, France\looseness=-1}%

\author{Tristan Meunier}
\affiliation{Univ. Grenoble Alpes, CNRS, Grenoble INP, Institut N\'eel, F-38000 Grenoble, France\looseness=-1}%


\begin{abstract}
    Recent demonstrations using electron spins stored in quantum dots array as qubits are promising for developing a scalable quantum computing platform.
    An ongoing effort is therefore aiming at the precise control of the quantum dots parameters in larger and larger arrays which represents a complex challenge.
    Partitioning of the system with the help of the inter-dot tunnel barriers can lead to a simplification for tuning and offers a protection against unwanted charge displacement.
    In a triple quantum dot system, we demonstrate  a nanosecond control of the inter-dot tunnel rate permitting to reach the two extreme regimes,
    large GHz tunnel coupling and sub-Hz isolation between adjacent dots.
    We use this novel development to isolate a sub part of the array while performing charge displacement and readout in the rest of the system.
    The degree of control over the tunnel coupling achieved in a unit cell should motivate future protocol development for tuning, manipulation and readout including this capability.
\end{abstract}

\maketitle


Arrays of quantum dots (QDs) are identified as one possible road for scaling up electron spin-based quantum processors \cite{vinetScalableSiliconQuantum2018,liCrossbarNetworkSilicon2018,veldhorstSiliconCMOSArchitecture2017,vandersypenInterfacingSpinQubits2017}.
In this context, the ability to displace controllably individual electrons plays an important role for realizing elementary operations within the array.
Displacement at the dot scale induces high fidelity coherent manipulation and interaction \cite{pettaCoherentManipulationCoupled2005b,brunnerTwoQubitGateCombined2011,watsonProgrammableTwoqubitQuantum2018,yonedaQuantumdotSpinQubit2018}.
Displacement at multi-dot scale enables array filling \cite{mortemousqueCoherentControlIndividual2021,volkLoadingQuantumdotBased2019}, and functionalities for long distance quantum interconnection \cite{mortemousqueCoherentControlIndividual2021,millsShuttlingSingleCharge2019a,jadotDistantSpinEntanglement2021}.
These displacement capabilities come with potential sources of errors such as incorrect positioning and tunneling while operating the electron spin qubits.
It is therefore desirable to find protocols to minimize their impact on the rest of the qubits.
The tunnel barrier control over a large range offers strategies to protect the electron spin information without losing quantum manipulation capabilities \cite{eeninkTunableCouplingIsolation2019}.
In semiconductor devices, this method is commonly used to isolate QD arrays from electron reservoirs, thereby fixing the total number of charges in the system \cite{bertrandQuantumManipulationTwoElectron2015,yangOperationSiliconQuantum2020a}.

Here we characterize the inter-dot tunnel rate from the sub-Hz to GHz regime and achieve complete isolation both from the reservoirs and the neighbor QD of up to three electrons.
Then, we implement two functionalities demonstrating the potential of the array partitioning process.
First, an improvement of metastable charge states lifetime and their readout at a fixed and optimized position, and then charge displacement and readout in the partitioned array.


\begin{figure}[]
    \centering
    \includegraphics[width=\singlecolumn]{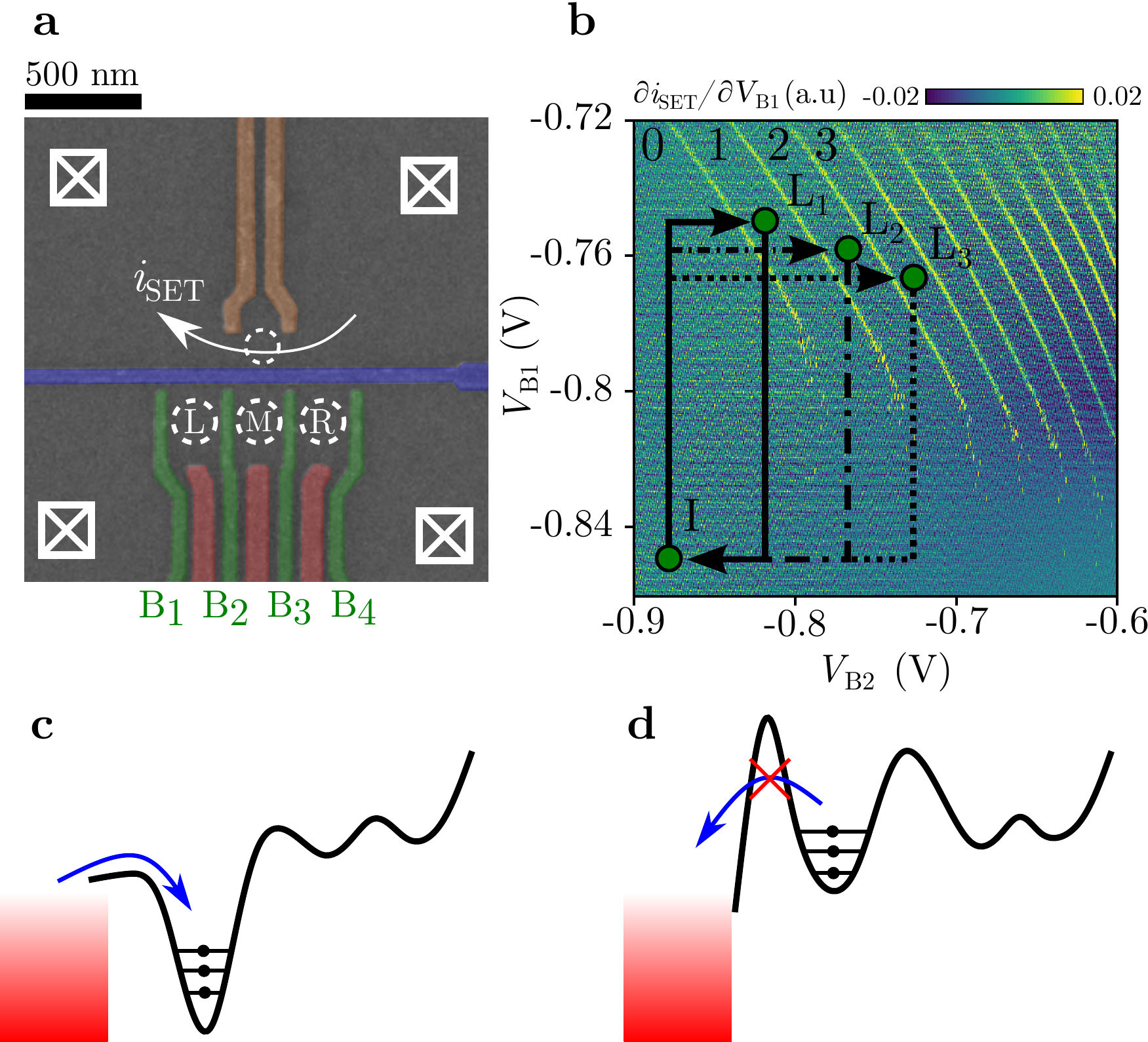}
    \caption{
        \textbf{Device, electron loading and isolation from the reservoirs.}
        \textbf{a,}~Electron micrograph of a sample similar to the one measured.
        \textbf{b,}~Stability diagram of the leftmost QD.
        The derivative of the current (with respect to $V_{\textrm{B1}}$)  measured across the SET is plotted as a function of the voltage applied on \Bone \ and \Btwo \ gates controlling respectively the reservoir-QD tunnel barrier height and the chemical potential of the QD.
        The electron occupation number is indicated for the first four charge regions of the stability diagram.
        The indicated positions $\textrm{L}_1$, $\textrm{L}_2$, $\textrm{L}_3$ and I are used to respectively load 1, 2, 3 electrons in the left QD and to isolate the QD array from the left electron reservoir.
        \textbf{c,}~Schematic of the potential landscape at position $\textrm{L}_3$ of the stability diagram.
        \textbf{d,}~Schematic of the potential landscape at position I of the stability diagram after the loading procedure.
    }
    \label{fig:FigSample}
\end{figure}

The device measured in this work is presented in Fig.~\ref{fig:FigSample}a and is composed of a linear triple QD array defined electrostatically by voltages applied to metallic gates on a GaAs/AlGaAs heterostructure.
The electrometer consists of a single electron transistor (SET) set on the side of a Coulomb peak to be sensitive to the charge configuration of the array.
The QD array is tuned by adjusting the voltage applied on the gates labelled as $\textrm{B}_{1-4}$.
The gate voltages applied for each experiment discussed are summarized in Sup. Mat.~{\SupVoltage}.
We first focus on the protocol to isolate a quantum dot from the reservoirs.
The first step is to load electrons from the bottom left reservoir (Fig.~\ref{fig:FigSample}a) to the previously emptied QD nanostructure.
We show in Fig.~\ref{fig:FigSample}b, a so-called stability diagram where we vary the voltages applied on \Bone \ and \Btwo \ while recording the current $i_{\textrm{SET}}$ through the electrometer.
In this diagram it is possible to determine the absolute number of electrons in the dot by identifying regions separated by charge degeneracy lines.
The chemical potential of L is controlled by the voltage applied on \Btwo \ while the reservoir to QD tunnel coupling is controlled by the voltage applied to \Bone, as indicated by the disappearance of the charge degeneracy lines for $V_{\textrm{B1}}<-0.8 \ \textrm{V}$.
The interruption of the degeneracy lines is an indicator of the isolated regime where the electron exchange rate with the reservoir is slower than the measurement sweep rate (250 mV/s).
We engineered a pulse sequence to load the desired number of electrons in the QD structure and isolate them from the reservoir.
The system is first initialized at point I empty of any electrons and the pulse sequence drawn on top of the stability diagram of Fig.~\ref{fig:FigSample}b is applied.
A voltage pulse on \Bone \ gate increases the coupling between the array and the reservoir, allowing the exchange of electrons with dot L.
Then the chemical potential of the left QD is lowered by applying a voltage pulse on gate \Btwo.
By varying the amplitude of this pulse to reach either the point \mbox{$\textrm{L}_1$, $\textrm{L}_2$ or $\textrm{L}_3$} it is possible to load respectively 1, 2 or 3 electrons in the dot.
A sketch of the potential landscape at the position $\textrm{L}_3$ is pictured in Fig.~\ref{fig:FigSample}c.
From the selected position, voltage pulses are applied on gate \Bone \ and then \Btwo \ to reach the position I, where the electron tunneling to the reservoir is suppressed (see Sup. Mat.~{\SupLoadingMap}).
In this configuration, the high chemical potential of the left QD guarantees that all loaded electrons should eventually tunnel back to the reservoir leaving the QD empty.
However, due to the low tunnel coupling to the reservoir this metastable configuration can be held for several tens of seconds (see Sup. Mat.~{\SupLoadingMap}).



\begin{figure*}[]
    \centering
    \includegraphics[width=\doublecolumn]{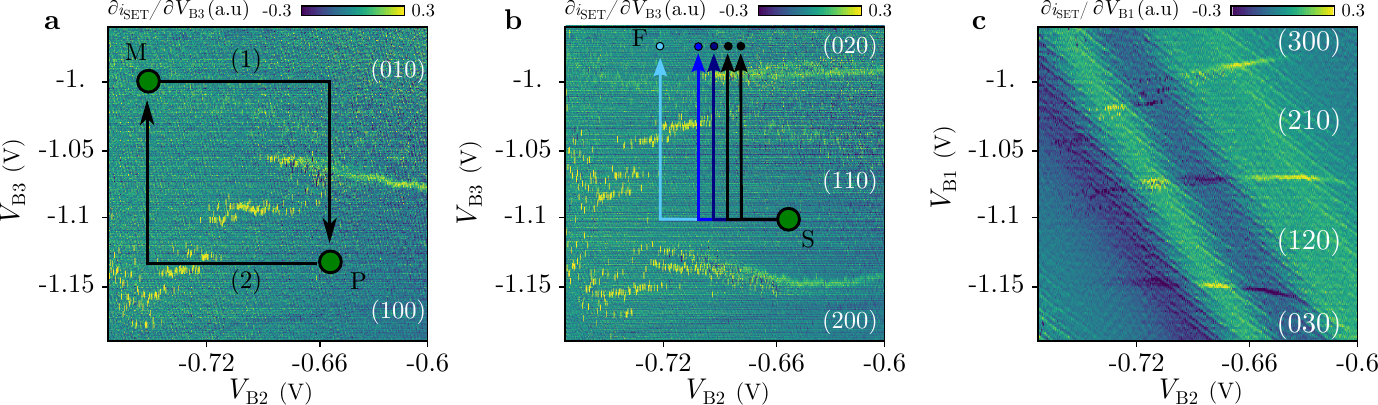}
    \caption{
        \textbf{1, 2 and 3 electrons stability diagram of a DQD.}
        \textbf{a,b,c,}~Stability diagram of the L-M DQD performed with a fixed number of charges respectively 1, 2 and 3 isolated from the reservoirs.
        The detuning and the tunnel coupling of the L-M DQD are swept using relevant gates \Bthree \ and \Btwo.
        It is possible to access all charge states of the DQD by sweeping \Bthree \ gate voltage over around 200 mV.
        For a negative enough voltage applied on \Btwo \ the stability diagram exhibits excited charge states of the DQD that are only observable in the low tunnel coupling regime.
        Charge configurations of the array are indicated in white.
    }
    \label{fig:StabDiag}
\end{figure*}

In this section, the isolation process is pushed one step further to perform array partitioning by demonstrating QD-to-QD decoupling.
We note the charge configuration of the array $(l,m,r)$ with $l$, $m$ and $r$ the charge occupation of the QDs L, M and R, respectively.
After loading either 1, 2 or 3 electrons in L, we vary the voltages applied on \Btwo \ and \Bthree \ to progressively transfer charges to dot M.
The corresponding stability diagrams are presented in \mbox{Fig. \ref{fig:StabDiag}a, b, c.}
For a system of $n$ dots containing a total of $k$ electrons, we expect $\binom{n+k-1}{k}$ charge states, which we experimentally observe for $n=2$ and $k$ up to 3.
Analogous to the QD-reservoir decoupling, we observe the apparition of stochastic events as $V_{\textrm{B2}}$ becomes increasingly more negative.
This phenomenon now corresponds to the L-M inter-dot tunnel rate becoming comparable to the measurement sweep rate (250 mV/s).

\begin{figure}[]
    \centering
    \includegraphics[width=\singlecolumn]{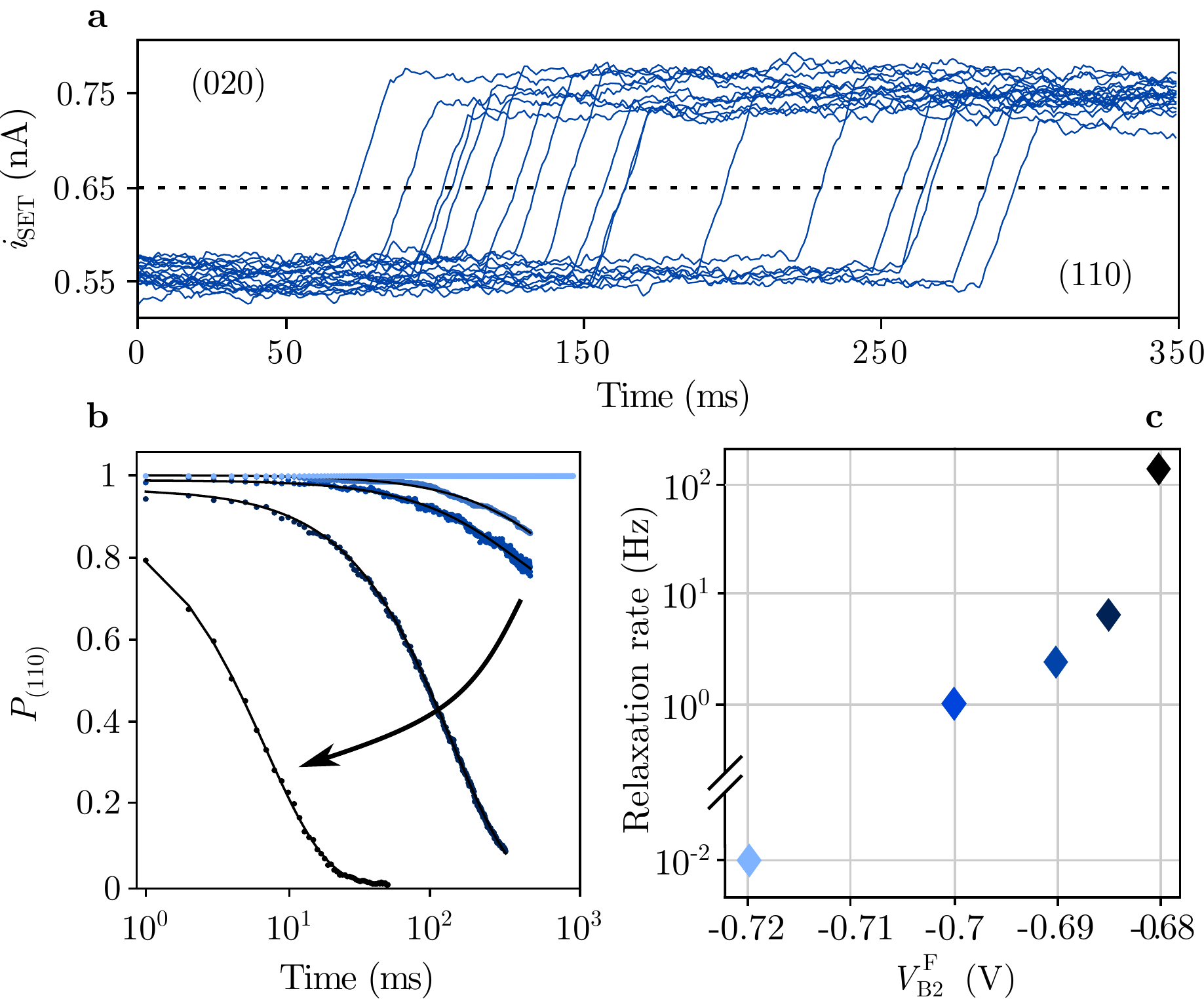}
    \caption{
        \textbf{Probing relaxation process of a metastable charge state in DQD.}
        \textbf{a,}~ Selected single shot measurements of $(110)$ to $(020)$ relaxation for $V_{\textrm{B2}}^{\textrm{F}}= -0.69 \ \textrm{V}$.
        The system is initialized to point P in the $(110)$ charge configuration.
        The tunnel barrier height between the dots is set by applying a pulse of varying amplitude $V_{\textrm{B2}}^{\textrm{F}}$.
        The system is then brought back in the region where the equilibrium charge state of the array is (020).
        The current $i_{\textrm{SET}}$ displays single event relaxation.
        A moving average filter is applied to the traces for clarity.
        \textbf{b,}~Relaxation of the (110) to (020) charge state observed for different tunnel barrier height.
        A threshold is defined halfway between the two current levels represented by a dashed line in \textbf{a}.
        For the \num{1000} traces, a current above or below this threshold is associated to respectively $(110)$ or $(020)$ charge state.
        The binarized traces are averaged for each $V_{\textrm{B2}}^{\textrm{F}}$ value tested to compute the $(110)$ population.
        Experimental data is represented as solid points and the solid black line is an exponential decay fit.
        For $V_{\textrm{B2}}^{\textrm{F}}= -0.72 \ \textrm{V}$ we do not observe any relaxation event in 1000 shots of 1 s.
        \textbf{c,}~Relaxation rate of the (110) to the (020) charge state as a function of the freeze point (F) coordinate.
    }
    \label{fig:Relaxation}
\end{figure}

In order to quantify the inter-dot tunnel rate dependence with the voltage applied on gate \Btwo, we designed the pulse sequence sketched on top of the stability diagrams in Fig.~\ref{fig:StabDiag}b.
Two electrons are loaded in L and the system is brought in the $(110)$ configuration at position S.
From this point, the tunnel coupling is lowered to the desired value $V_{\textrm{B2}}^{\textrm{F}}$ using a voltage pulse on the gate \Btwo.
After 100 ns the detuning is set to reach the $(020)$ charge region via a voltage pulse on gate \Bthree.
At this position the $(110)$ charge state becomes metastable.
To track the evolution of the charge state, we record the current $i_{\textrm{SET}}$ during up to 1 s.
The procedure is repeated 1000 times for $V_{\textrm{B2}}^{\textrm{F}}$ between -0.72 V and -0.68 V.
Selected records of $i_{\textrm{SET}}$ for $V_{\textrm{B2}}^{\textrm{F}} = -0.69 \ \textrm{V}$ are shown in Fig.~\ref{fig:Relaxation}a and we observe sharp single jumps of $i_{\textrm{SET}}$ from 0.55 nA to 0.75 nA.
These events are associated to an electron tunneling from M to L.
In Fig.~\ref{fig:Relaxation}b we compute the probability $P_{(110)}$ to observe the $(110)$ charge state as a function of the waiting time at point F and observe an exponential decay of the population.
In Fig.~\ref{fig:Relaxation}c, we observe that the charge state lifetime can be tuned over 4 orders of magnitude in few tens of mV.
In particular, for $V_{\textrm{B2}}^{\textrm{F}}\leqslant -0.72 V$, no relaxation events are visible in a thousand 1 s-long time-traces, setting a higher tunnel rate bound at $10^{-2}$ Hz.
This demonstrates that we are able to reduce the inter-dot tunnel rate well below the Hz regime while keeping the initial QD structure intact.
Moreover, the high level of control in the low inter-dot tunnel coupling regime did not prevent us to perform spin qubit operation which requires a GHz tunnel coupling in the same sample with the same tuning (data not shown here).

\begin{figure*}[]
    \centering
    \includegraphics[width=\doublecolumn]{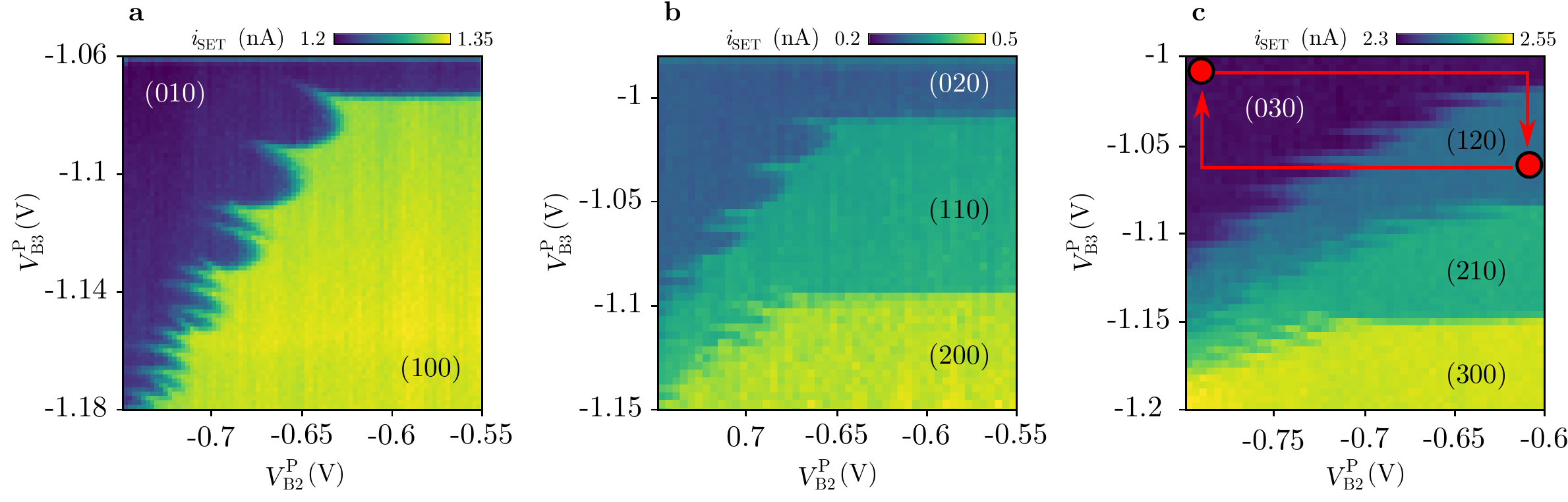}
    \caption{
        \textbf{1, 2 and 3 electrons freeze map in a DQD.}
        \textbf{a,b,c,}~1, 2 and 3 electrons freeze map of the L-M DQD.
        The measurements are obtained by performing the pulse sequence sketched on top of the stability diagram in Fig.~\ref{fig:StabDiag}a.
        The first pulse probes the DQD at a certain value of detuning and tunnel coupling at a varied point P.
        The second one brings back the system at point M by first setting the tunnel coupling in the sub-Hz regime and then the detuning.
        Finally, the current $i_{\textrm{SET}}$ is recorded during 5 ms.
    }
    \label{fig:FreezeMap}
\end{figure*}

The capability to operate over such a wide range the inter-dot tunnel coupling enables novel functionalities for future prospects in spin qubit technology \cite{liCrossbarNetworkSilicon2018}.
Indeed, freezing on a fast timescale the electron dynamics results in a well separated and metastable charge configuration that can be efficiently probed.
Proof of principle experiment is performed in a tunnel coupled double quantum dot (DQD) with up to three electrons.
The protocol consists in loading a specific charge configuration in the double dot, decrease on fast timescale the tunnel barrier and then tune the system to a working point at which the charge detection has been optimized while preserving the charge configuration.

To do so we manipulate the inter-dot tunnel coupling and the detuning of the L-M DQD at the nanosecond timescale.
The initialization of metastable charge configuration of the array is characterized using a freeze map protocol.
It consists in setting the system at a given detuning and tunnel coupling value before pulsing the inter-dot tunnel rate to the sub-Hz regime to freeze the charge configuration.
Followed by a charge readout, this protocol allows us to identify the detuning and tunnel coupling regions where a charge transfer is possible.

In addition to the already described notation for labelling the charge states we introduce a vertical bar $\lvert$ indicating a sub-Hz tunnel coupling rate in between the QDs.
The trajectory, visible in Fig.~\ref{fig:StabDiag}a, starts in the $(0\lvert10)$ charge state at point M and is used to realize the freeze map protocol.
Two 100 ns voltages pulses applied sequentially to \Btwo \ and \Bthree \ set the system to point P.
Then, the inter-dot tunnel rate is lowered to the sub-\si{\hertz} regime and the detuning is set back to position M.
At this position the electrometer signal $i_{\textrm{SET}}$ is averaged during 5 ms.
Depending on the coordinates of point P($V_{\textrm{B2}}^{\textrm{P}}$,$V_{\textrm{B3}}^{\textrm{P}}$), we obtain two possible values for $i_{\textrm{SET}}$ corresponding to either the $(0\lvert10)$ or $(1\lvert00)$ charge state, as shown in Fig.~\ref{fig:FreezeMap}a.
Comparing with the stability diagram in Fig.~\ref{fig:StabDiag}a, for $V_{\textrm{B2}} > -0.65 \ \textrm{V}$ we observe the same charge transition at $V_{\textrm{B3}} = -1.07 \ \textrm{V}$.
It indicates that the time spent at P is sufficient to observe charge transfer.
For $V_{\textrm{B2}} < -0.65 \ \textrm{V}$, the tunnel coupling rate is not strong enough to allow charge transfer in the ground state at $V_{\textrm{B3}} = -1.07 \ \textrm{V}$.
Nevertheless, by increasing the detuning ($V_{\textrm{B3}}< -1.07 \ \textrm{V}$) higher orbital states of M with higher tunnel coupling are accessible and allow the charge transfer within the time spent in P.
The complete description of the charge dynamics to account for the precise shape of the charge transition is beyond the scope of this work and will be studied further in subsequent experiments.
The agreement between the freeze map and the stability diagram is also observed for 2 and 3 electrons loaded in the array as shown in Fig.~\ref{fig:FreezeMap}b,c.
To conclude, we demonstrated the capability to initialize metastable charge configuration for a duration long enough to permit their readout at an optimized position in the voltage gate space.
This study is a first demonstration of the novel initialization and readout protocols induced by the high level of control over the inter-dot tunnel coupling.

\begin{figure}[]
    \centering
    \includegraphics[width=\singlecolumn]{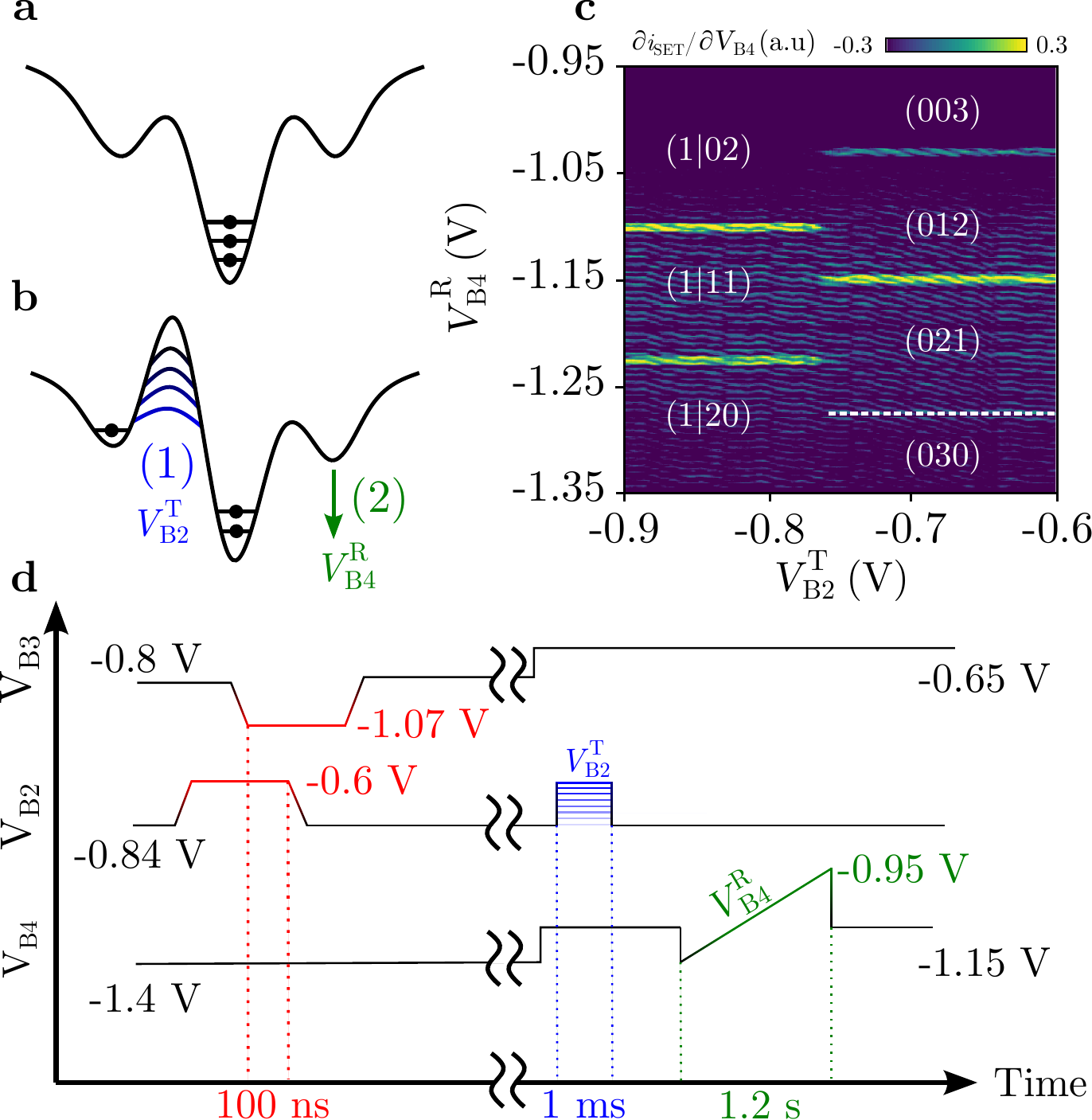}
    \caption{
        \textbf{Isolating sub parts of the linear QD array.}
        \textbf{a,}~Schematic of the potential landscape at the initialization point.
        \textbf{b,}~Schematic of the potential landscape at the beginning of the stability diagram of figure \textbf{c}.
        \textbf{c,}~Stability diagram performed while a a part of the array is segmented.
        For $V^{\textrm{T}}_{\textrm{B2}} <-0.75 \ \textrm{V}$ the electron in L is unable to tunnel back to M and the only charge state seen on the stability diagram are ($1\vert20$), ($1\vert11$) and ($1\vert02$).
        For higher values of $V^{\textrm{T}}_{\textrm{B2}}$ the electron in L tunnels back to M and all charge states of 3 electrons in a DQD are observed.
        \textbf{d,}~Chronograph of the voltages applied on gate \Btwo, \Bthree \ and \Bfour.
        The pre-sequence is dedicated to create a metastable charge state ($1\vert20$) using the pulse sequence described in Fig.~\ref{fig:FreezeMap}c.
        Then the voltage applied on \Btwo \ gate is pulsed during 1 ms to $V^{\textrm{T}}_{\textrm{B2}}$ and finally the chemical potential of the R QD is ramped while the current $i_{\textrm{SET}}$ is recorded.
    }
    \label{fig:PartialFreeze}
\end{figure}

In addition to an improvement of the charge determination in QD arrays, the inter-dot tunnel coupling control also grants us the possibility to isolate subparts of a QD array to simplify the tuning and manipulation.
In this section we demonstrate the complete isolation of a sub part of the QD array while keeping the complete control over the charge configuration in the rest of the system.
To do so we initialize a metastable charge state of the L-M DQD in order to isolate an electron in L.
Then we progressively transfer the charges remaining in M to R to control the number of electrons present in the M-R DQD subsystem.

The first step to implement this protocol consists in loading 3 electrons in M to reach the (030) charge state.
From there, a ($1\vert20$) metastable charge configuration is initialized by pulsing the voltages applied on gate \Btwo \ and \Bthree, an equivalent of the pulse sequence performed is sketched on top of the freeze map in Fig.~\ref{fig:FreezeMap}c.
The next step consists in opening the tunneling between M and R by applying -0.65 V on gate \Bthree, and lowering the chemical potential of R closer to the one of M by increasing the voltage applied on \Bfour \ to -1.15 V.
In this voltage configuration the L-M inter-dot tunnel coupling is pulsed during 1 ms using a voltage pulse of amplitude $V^{\textrm{T}}_{\textrm{B2}}$.
Following the pulse, the detuning of L-R is ramped using the voltage applied on gate \Bfour \ while $i_{\textrm{SET}}$ is recorded.
The derivative $\partial i_{\textrm{SET}}/\partial V_{\textrm{B4}}$ is plotted as a function of $V_{\textrm{B4}}^{\textrm{R}}$ and $V_{\textrm{B2}}^{\textrm{T}}$ in Fig.~\ref{fig:PartialFreeze}c.
For $V_{\textrm{B2}}^{\textrm{T}}<-0.75 \ \textrm{V}$ we observe two degeneracy lines in the stability diagram indicating that the sub array composed of M and R QDs contains only two electrons while the third one is isolated in L.
Due to the low L-M inter-dot tunnel coupling, the electron in L cannot tunnel back to M during the whole 1.2 s ramp, in this configuration the only charge states available by the array are $(1\lvert02)$, $(1\lvert11)$ and $(1\lvert02)$ (see Sup. Mat.~{\SupManipSubPart}).
They are identified and labelled on top of the stability diagram.
For a pulse amplitude $V_{\textrm{B2}}^{\textrm{T}}> -0.75 \ \textrm{V}$, we observe a third line, indicating that the electron stored in L tunneled back to M during the $V_{\textrm{B2}}^{\textrm{T}}$ pulse.
Indeed, this configuration allows the relaxation of the (120) to the (030) charge state and the resulting stability diagram corresponds to a classical one for three electrons loaded in a DQD.
To conclude, we are able to perform electron manipulation in a partitioned DQD while preserving the charge state in the adjacent QD.
This study demonstrates the ability of the array partitioning to lower the complexity of the stability diagrams by reducing the number of charge states available for the electrons.



The control of the tunnel couplings and the chemical potential of each QD on fast timescales allowed us to initialize arbitrary metastable charge state of up to three electrons in a DQD.
The freeze map protocol developed in this article allowed us to enhance the lifetime and perform readout of these states at a fixed and optimized position in the voltage gate space.
This demonstration is of particular interest in the context of the wide use of Pauli spin blocked spin to charge conversion whose fidelity is limited by the lifetime of such metastable charge states \cite{barthelRapidSingleShotMeasurement2009}.
We finally performed a segmentation of the array by decoupling a QD filled with one electron while performing charge displacement and readout in the rest of the structure.
Doing so we observed a reduction of the charge states available for the system and therefore a reduction of the complexity while tuning the QD array.
The partitioning protocol opens the door to more complex applications such as the operation of larger 1D or 2D arrays of QDs while keeping the low dimensionality of simple sub systems \cite{millsComputerautomatedTuningProcedures2019}.


See supplementary material for additional information of the array tuning and the array partitioning protocol.


We acknowledge technical support from the Pole groups of the Institut Néel, and in particular the NANOFAB team who helped with the sample realization, as well as T. Crozes, E. Eyraud, D. Lepoittevin, C. Hoarau and C. Guttin.
This work is supported by the ERC QUCUBE.
A.L. and A.D.W. acknowledge gratefully support of DFG-TRR160 and DFH/UFA CDFA-05-06, DFG project 383065199, and BMBF QR.X Project 16KISQ009.

The data that support the findings of this study are available from the corresponding author upon reasonable request.


The authors have no conflicts to disclose.

M.N. and V.T. fabricated the sample.
M.N. performed the experiment with the help of B.J.
M.N. interpreted the data with the help of B.J., P.-A.M., E.C., M.D. and T.M.
A.L. and A.D.W. performed the design and molecular-beam-epitaxy growth of the high-mobility heterostructure.
All the authors discussed the results extensively, as well as the manuscript.

\bibliography{biblio}

\end{document}



\title{Supplementary Material for: Controlled quantum dot array segmentation via a highly tunable interdot tunnel coupling}

\author{Martin Nurizzo}
\affiliation{Univ. Grenoble Alpes, CNRS, Grenoble INP, Institut N\'eel, F-38000 Grenoble, France\looseness=-1}%

\author{Baptiste Jadot}
\affiliation{Univ. Grenoble Alpes, CEA, Leti, F-38000 Grenoble, France}%

\author{Pierre-Andr\'e Mortemousque}
\affiliation{Univ. Grenoble Alpes, CEA, Leti, F-38000 Grenoble, France}%

\author{Vivien Thiney}
\affiliation{Univ. Grenoble Alpes, CEA, Leti, F-38000 Grenoble, France}%

\author{Emmanuel Chanrion}
\affiliation{Univ. Grenoble Alpes, CNRS, Grenoble INP, Institut N\'eel, F-38000 Grenoble, France\looseness=-1}%

\author{Matthieu Dartiailh}
\affiliation{Univ. Grenoble Alpes, CNRS, Grenoble INP, Institut N\'eel, F-38000 Grenoble, France\looseness=-1}%

\author{Arne Ludwig}
\affiliation{Lehrstuhl f{\"u}r Angewandte Festk{\"o}rperphysik, Ruhr-Universit{\"a}t Bochum, Universit{\"a}tsstra{\ss}e 150, D-44780 Bochum, Germany}%

\author{Andreas D. Wieck}
\affiliation{Lehrstuhl f{\"u}r Angewandte Festk{\"o}rperphysik, Ruhr-Universit{\"a}t Bochum, Universit{\"a}tsstra{\ss}e 150, D-44780 Bochum, Germany}%

\author{Christopher B{\"a}uerle}
\affiliation{Univ. Grenoble Alpes, CNRS, Grenoble INP, Institut N\'eel, F-38000 Grenoble, France\looseness=-1}%

\author{Matias Urdampilleta}
\affiliation{Univ. Grenoble Alpes, CNRS, Grenoble INP, Institut N\'eel, F-38000 Grenoble, France\looseness=-1}%

\author{Tristan Meunier}
\affiliation{Univ. Grenoble Alpes, CNRS, Grenoble INP, Institut N\'eel, F-38000 Grenoble, France\looseness=-1}%

\maketitle
\newpage


\renewcommand{\theequation}{S\arabic{equation}}
\renewcommand{\thefigure}{S\arabic{figure}}
\renewcommand{\bibnumfmt}[1]{[#1]}
\renewcommand{\citenumfont}[1]{#1}

\newcommand{\ket}[1]{\left| #1 \right>} 
\newcommand{\bra}[1]{\left< #1 \right|} 
\newcommand{\matrixel}[3]{\big< #1 \big| #2 \big| #3\big>} 
\newcommand{\spinT}[2]{\left| #1 \right> \leftrightarrow \left| #2\right>} 
\newcommand{\scalarprod}[2]{\big< #1\big| #2\big>} 
\newcommand{\Bone}{$\textrm{B}_{\textrm{1}}$}
\newcommand{\Btwo}{$\textrm{B}_{\textrm{2}}$}
\newcommand{\Bthree}{$\textrm{B}_{\textrm{3}}$}
\newcommand{\Bfour}{$\textrm{B}_{\textrm{4}}$}

\externaldocument{charge_freezing_main}

\section{Materials and set-up}

Our device was fabricated using a Si-doped AlGaAs/GaAs heterostructure grown by molecular beam epitaxy, with a two-dimensional electron gas 100 nm below the crystal surface which has a carrier mobility of $9\times10^5$ cm$^2$V$^{-1}$s$^{-1}$ and an electron density of $2.7 \times 10^{11}$ cm$^2$.
The device is micro-bonded to a homemade PCB providing DC and RF lines with cut-off frequencies of respectively 10 MHz and 10 GHz.
It is thermally anchored to the mixing chamber of a dry dilution refrigerator with a base temperature of 15 mK.
QDs are defined and controlled by the application of negative voltages on Ti/Au Schottky gates deposited on the surface of the crystal.
A homemade DAC ensures fast changes of both chemical potentials and tunnel couplings with voltage pulse rise times approaching 100 ns and refreshed every 6 $\mu$s.
A Tektronix 5014C AWG with a typical channel voltage rise time (20\% - 80\%) of 0.9 ns is used to change the $V_\textrm{B2}$ and $V_\textrm{B3}$ gate voltages at the nanosecond timescale.
These fast gates are connected to -17 dB attenuated RF lines and the signal is added to the DAC signal through a homemade bias-tee with a cut-off frequency of 1 kHz.
The sum of the two signals is computed by assuming that 1 V applied on the AWG corresponds to approximately 0.1 V on the sample.
The gate lever arm is expected to be $\alpha \simeq 0.05 \ \textrm{eV}.\textrm{V}^{-1}$ in concordance with previously measured devices.
The charge configurations can be read out by a QD tuned to be a sensitive local electrometer and biased with 100 $\mu$V.

\section{Loading the array in the isolated regime}

\begin{figure}[ht]
    \centering
    \includegraphics[width=\doublecolumn]{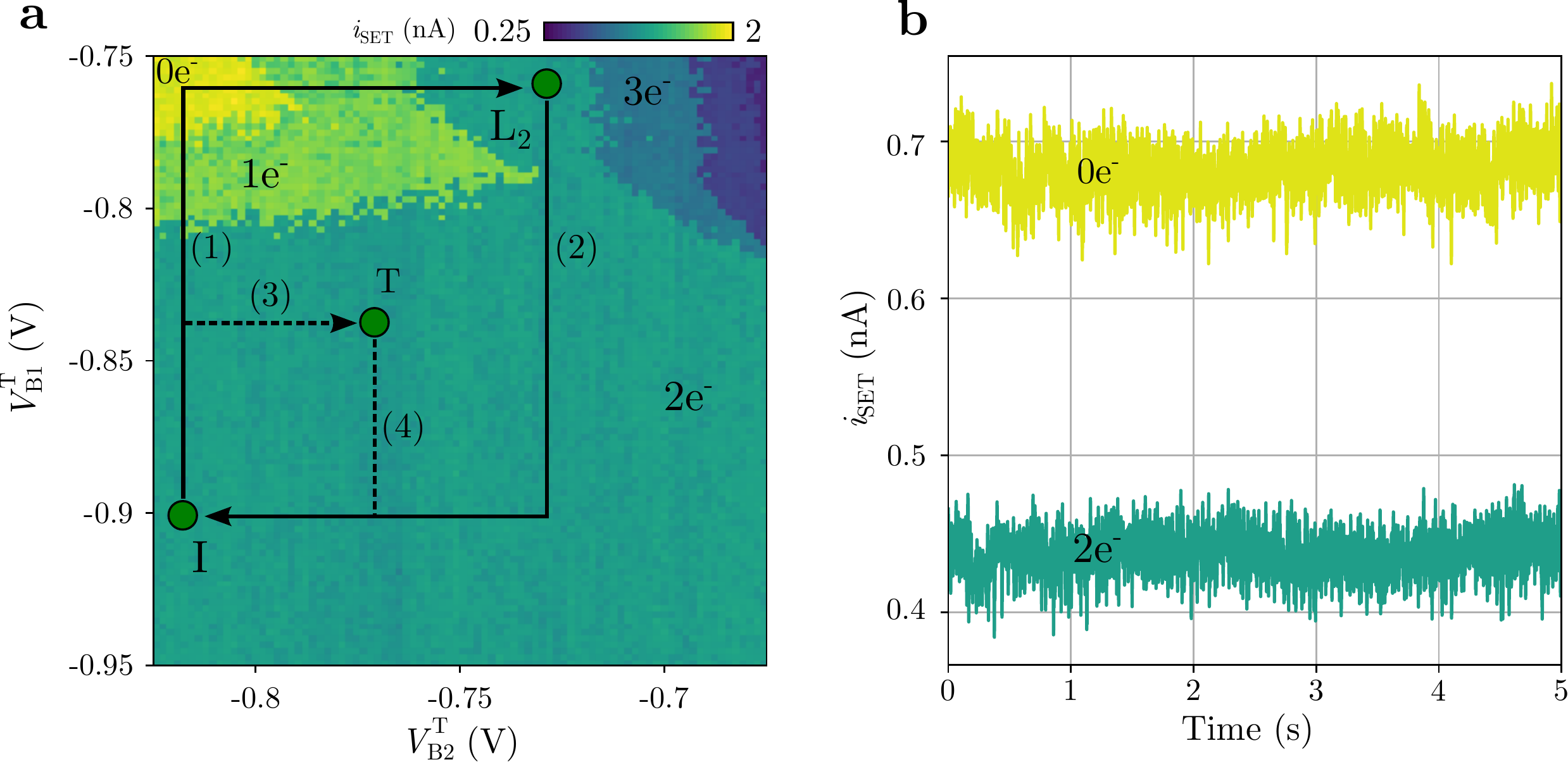}
    \caption{
    \textbf{Electron loading procedure and limits of the isolated regime in the gate voltage space.}
    \textbf{a,}~Losing map of L identifying the limits of the isolated regime.
    Each pixel corresponds to an experiment shot where two electrons are loaded in L and isolated from the reservoirs using pulse (1) and (2).
    The system is then pulsed to the varied position T during 1 ms using pulse (3) and brought back to I using pulse (4).
    Finally, the current $\textrm{i}_{\textrm{SET}}$ is averaged during 5 ms.
    \textbf{b,}~QD charge occupation at position I after initially loading zero or two electrons.
    No electron tunneling in or out the dot events were recorded for the 5 s-long 1000 shots.
    \label{fig:SUP-losing-map}
    }
\end{figure}

The loading procedure of the electrons in the isolated regime of the array is calibrated using the experiment described in Fig.~\ref{fig:SUP-losing-map}a.
The array is initialized empty of any electrons at position I, then the reservoir-QD tunnel coupling is increased via a voltage pulse  on \Bone \ and the chemical potential of the QD is lowered using a voltage pulse on \Btwo \ to reach position $\textrm{L}_2$.
Once two electrons are loaded in L the QD is isolated from the reservoirs by bringing the system back to position I.
We now want to identify the region in the voltage gate space where the two previously loaded electrons are kept in the array to define the isolated regime.
To do so, the system is pulsed to a varied point T during 1 ms before being set back to position I to perform a charge readout by acquiring $\textrm{i}_{\textrm{SET}}$ during 5 ms.
In Fig.~\ref{fig:SUP-losing-map}a we plotted the averaged current as a function of the coordinates of point T($\textrm{V}_{B2}^T$,$\textrm{V}_{B1}^T$).
We identify and label different levels of $\textrm{i}_{\textrm{SET}}$ corresponding to different charge occupations of L.
As expected for more negative voltages on \Bone \ we observe a large region where the two initially loaded electrons are kept in the array whatever the chemical potential of the QD.
This region is identified as the isolated regime, indeed here the reservoir-QD tunnel coupling is low enough to prevent any tunneling events from or to the reservoirs.
For $V_{\textrm{B1}}^{\textrm{T}}> -0.8 \ \textrm{V}$ the exchange of electrons is again possible and we observe the ability to reduce the occupation the QD for $V_{\textrm{B2}}^{\textrm{T}}< -0.75 \ \textrm{V}$ and to increase it for $V_{\textrm{B2}}^{\textrm{T}}> -0.75 \ \textrm{V}$.

Now that the isolated regime region is identified we want to ensure the capability to isolate the electrons from the reservoirs during a time longer than the typical experiment (1 s).
We either load zero or two electrons in L and isolate the QD from the reservoirs at position I in Fig.~\ref{fig:SUP-losing-map}a using the same pulse sequence described before.
The current $i_{\textrm{SET}}$ is acquired during 5 s and the result is plotted in Fig.~\ref{fig:SUP-losing-map}b, a variation of the charge occupation in the isolated regime would result in a jump of the SET current.
As expected, we observe no such events in 5 s and to ensure this observation we repeated the experiment 1000 times and never observed a charge variation in the array.
The probability to have no relaxation events in 1000 experiments of 5 s is defined as $e^{-5000/\tau}$ whre $\tau$ is the characteristic isolation time.
We can therefore ensure, with a 99\% probability, that $\tau> 1100 \ \textrm{s}$. 

\section{Charge manipulation in the fully isolated regime}
\begin{figure*}[ht]
    \centering
    \includegraphics[width=\doublecolumn]{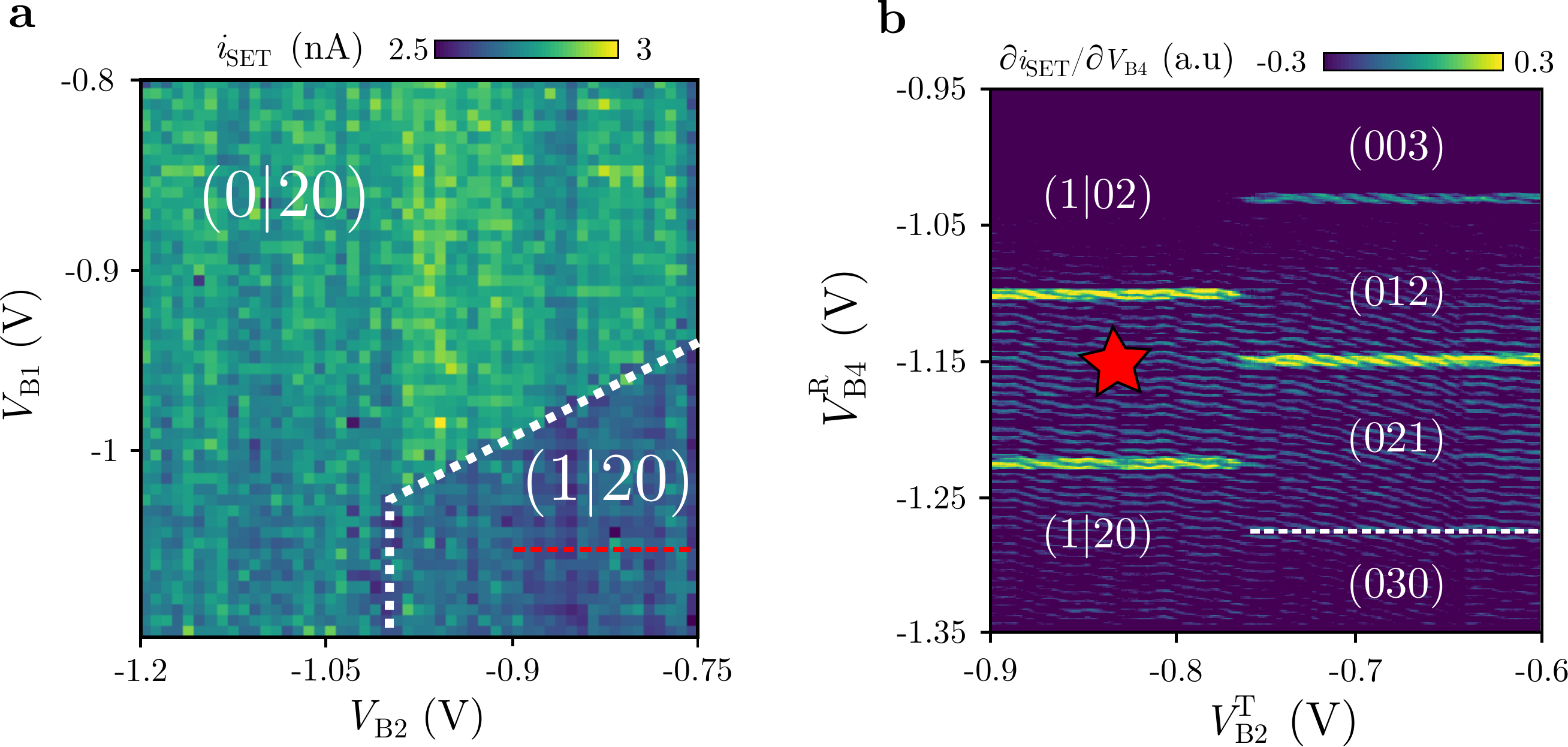}
    \caption{
        \textbf{Checking the charge state of L during the partial freeze.}
        \textbf{a,}~Charge detection in L after a partial freeze of the array in the $(1\lvert 11)$ charge configuration.
        The system is pulsed at the red star in Fig.~\ref{fig:SUP-PartialFreeze} to create a $(1\lvert 11)$ configuration during 1 ms before coming back in the $(1\lvert 20)$ region.
        At this position we pulse the voltage applied on gate \Bone \ and \Btwo \ during 1 ms and we record the electrometer signal.
        Two regions of different signal level are observed discriminating two charge configuration of the array.
        For high values of voltage applied on \Bone \ the tunneling to the left reservoir becomes possible and the system is no more in the isolated regime.
        We therefore label this region as $(0\lvert 20)$ and the second one to $(1\lvert 20)$.
        A dashed red line is here to represent the axis on which the array segmentation experiment is performed Fig.~5.
        The ability to observe an electron exiting the system from L at the end of the partial freeze experiment demonstrates the presence of all the electrons during the whole experiment.
    }
    \label{fig:SUP-PartialFreeze}
\end{figure*}

The observations made on Fig.~5 are interpreted as the ability to set the array in a configuration where one electron is present in each QD while the left electron is no more coupled to the rest of the array $(1\lvert11)$.
However, it is also possible to explain this data by assuming that the left electron has been lost to the left reservoir during the manipulation of the tunnel barrier.
In Fig.~\ref{fig:SUP-PartialFreeze}, we ensure that the array still contains three electrons at the end of the experiment by emptying L in to reach the $(0\lvert20)$ configuration.
To do so we perform the same manipulation presented in Fig.~5 to initialize the array in the $(1\lvert11)$ configuration, this position is held during 1 ms before coming back to $(1\lvert20)$.
We then verify the presence of an electron in L by pulsing the voltage on \Bone \ and \Btwo \ during 1 ms on a large range before performing a charge readout.
We plot in Fig.~\ref{fig:SUP-PartialFreeze}a the output of the electrometer as a function of the voltage pulses amplitude and we observe two regions delimited by the white dashed line.
This variation of the charge configuration is associated to the electron in L exiting the array to left reservoir.
We can therefore label the two regions as $(0\lvert20)$ for the one at high chemical potential and high QD-reservoir coupling and $(1\lvert20)$ for the other one.
We finally ensure that the experiment presented in Fig.~5 is performed in the $(1\lvert20)$ region proving that no electrons are lost during the manipulation.

\section{Array tuning and applied voltages}

\begin{table*}[ht]
    \begin{ruledtabular}
        \begin{tabular}{ccccccccc}
                            & \textbf{B1}    & \textbf{T1} & \textbf{B2}    & \textbf{T2} & \textbf{B3}    & \textbf{T3} & \textbf{B4}    & \textbf{H} \\\hline
            \textbf{Fig 1b} & -0.72 to -0.85 & -0.55       & -0.9 to -0.6   & -0.85       & -0.9           & 0           & 0              & -1         \\
            \textbf{Fig 2a} & -1.1           & -0.55       & -0.8 to -0.6   & -0.85       & -1.01 to -1.19 & 0           & 0              & -1         \\
            \textbf{Fig 2b} & -1.1           & -0.55       & -0.8 to -0.6   & -0.85       & -1.01 to -1.19 & 0           & 0              & -1         \\
            \textbf{Fig 2c} & -0.95 to -1.2  & -0.55       & -0.8 to -0.6   & -0.85       & -0.95          & -0.55       & -1             & -1         \\
            \textbf{Fig 3a} & -1.1           & -0.55       & -0.75 to -0.55 & -0.85       & -1.06 to -1.18 & 0           & 0              & -1         \\
            \textbf{Fig 3b} & -1.1           & -0.55       & -0.75 to -0.55 & -0.85       & -0.98 to -1.15 & 0           & 0              & -1         \\
            \textbf{Fig 3c} & -1.1           & -0.55       & -0.8 to -0.6   & -0.45       & -1 to -1.2     & -0.55       & -1.4           & -1         \\
            \textbf{Fig 4b} & -1.1           & -0.55       & -0.9 to -0.6   & -0.45       & -1.            & -0.55       & -0.95 to -1.35 & -1         \\
        \end{tabular}
    \end{ruledtabular}
    \caption{Voltages applied on the gates of the device to perform all the experiments presented.}
    \label{table-voltage}
\end{table*}
